\documentclass[%
reprint,
superscriptaddress,
amsmath,amssymb,
pra,
]{revtex4-1}

\usepackage{graphicx}
\usepackage{dcolumn}
\usepackage{bm}
\usepackage{soul}
\usepackage[centerlast]{caption}
\usepackage{float}
\usepackage{tabularx}
\usepackage{array}
\usepackage{multirow}
\newcolumntype{Y}{>{\centering\arraybackslash}X}
\usepackage{siunitx}
\usepackage{hyperref}
\hypersetup{colorlinks=true, citecolor=blue, urlcolor=blue, linkcolor=blue}
\usepackage{todonotes}
\usepackage{xcolor}
\usepackage{colortbl}
\makeatletter
    \def\CT@@do@color{%
      \global\let\CT@do@color\relax
            \@tempdima\wd\z@
            \advance\@tempdima\@tempdimb
            \advance\@tempdima\@tempdimc
    \advance\@tempdimb\tabcolsep
    \advance\@tempdimc\tabcolsep
    \advance\@tempdima2\tabcolsep
            \kern-\@tempdimb
            \leaders\vrule
                    \hskip\@tempdima\@plus  1fill
            \kern-\@tempdimc
            \hskip-\wd\z@ \@plus -1fill }
    \makeatother
\usepackage{colortbl}

\begin{document}

\title[Depairing Current]%
{Determining the depairing current in superconducting \\ nanowire single-photon detectors}

\author{S. Frasca}
\email[E-mail: ]{simone.frasca@epfl.ch}
\altaffiliation[Now at: ]{Advanced Quantum Architecture Laboratory (AQUA), \'{E}cole Polytechnique F\'{e}d\'{e}rale de Lausanne at Microcity, 2002 Neuch\^{a}tel, Switzerland.}
\affiliation{Jet Propulsion Laboratory, California Institute of Technology, 4800 Oak Grove Dr., Pasadena, California 91109, USA}
 
\author{B. Korzh}
\email[E-mail: ]{bkorzh@jpl.caltech.edu}
\affiliation{Jet Propulsion Laboratory, California Institute of Technology, 4800 Oak Grove Dr., Pasadena, California 91109, USA}

\author{M. Colangelo}
\affiliation{Department of Electrical Engineering and Computer Science, Massachusetts Institute of Technology, Cambridge, Massachusetts 02139, USA}

\author{D. Zhu}
\affiliation{Department of Electrical Engineering and Computer Science, Massachusetts Institute of Technology, Cambridge, Massachusetts 02139, USA}

\author{A. E. Lita}
\affiliation{National Institute of Standards and Technology, Boulder, Colorado 80305, USA}

\author{J. P. Allmaras}
\affiliation{Jet Propulsion Laboratory, California Institute of Technology, 4800 Oak Grove Dr., Pasadena, California 91109, USA}
\affiliation{Department of Applied Physics, California Institute of Technology, Pasadena, California 91109, USA}

\author{E. E. Wollman}
\affiliation{Jet Propulsion Laboratory, California Institute of Technology, 4800 Oak Grove Dr., Pasadena, California 91109, USA}

\author{V. B. Verma}
\affiliation{National Institute of Standards and Technology, Boulder, Colorado 80305, USA}

\author{A. E. Dane}
\affiliation{Department of Electrical Engineering and Computer Science, Massachusetts Institute of Technology, Cambridge, Massachusetts 02139, USA}

\author{E. Ramirez}
\affiliation{Jet Propulsion Laboratory, California Institute of Technology, 4800 Oak Grove Dr., Pasadena, California 91109, USA}

\author{A. D. Beyer}
\affiliation{Jet Propulsion Laboratory, California Institute of Technology, 4800 Oak Grove Dr., Pasadena, California 91109, USA}

\author{S. W. Nam}
\affiliation{National Institute of Standards and Technology, Boulder, Colorado 80305, USA}

\author{A. G. Kozorezov}
\affiliation{Department of Physics, Lancaster University, Lancaster, UK, LA1 4YB}

\author{M. D. Shaw}
\affiliation{Jet Propulsion Laboratory, California Institute of Technology, 4800 Oak Grove Dr., Pasadena, California 91109, USA}

\author{K. K. Berggren}
\affiliation{Department of Electrical Engineering and Computer Science, Massachusetts Institute of Technology, Cambridge, Massachusetts 02139, USA}
\date{\today}

\begin{abstract}
\quad We estimate the depairing current of superconducting nanowire single-photon detectors (SNSPDs) by studying the dependence of the nanowires' kinetic inductance on their bias current. The kinetic inductance is determined by measuring the resonance frequency of resonator-style nanowire coplanar waveguides both in transmission and reflection configurations. Bias current dependent shifts in the measured resonant frequency correspond to the change in the kinetic inductance, which can be compared with theoretical predictions. We demonstrate that the fast relaxation model described in the literature accurately matches our experimental data and provides a valuable tool for direct determination of the depairing current. Accurate and direct measurement of the depairing current is critical for nanowire quality analysis, as well as modeling efforts aimed at understanding the detection mechanism in SNSPDs.
\end{abstract}

\maketitle

%

\section{\label{sec:intro}Introduction}

\quad Superconducting nanowire single-photon detectors (SNSPDs) \cite{goltsman_picosecond_2001} are established as a key technology for many applications, such as deep-space optical communication, laser ranging and quantum  science. This is due to their high efficiency ($>$ 90\%) \cite{Marsili_effic_2013}, wide wavelength sensitivity (from X-rays to mid-infrared) \cite{zhang_xsnspd, Kornev_midir_2012}, low dark count rate ($<$ 1 Hz) \cite{Wollman_17} and ultra-high timing resolution ($<$ 3~ps) \cite{korzh_demonstrating_2018}.

\quad In the last decade, widespread effort by the SNSPD community has improved the theoretical understanding of the detection mechanism in SNSPDs. Guided by experimental measurements \cite{Kerman_Constriction_2007, engel_temperature-dependence_2013, renema_experimental_2014, lusche_effect_2014, renema_effect_2015, marsili_hotspot_2016, gaudio_experimental_2016, caloz_optically_2017} and theoretical modeling \cite{bulaevskii_vortex-assisted_2012, vodolazov_vortex-assisted_2015, kozorezov_quasiparticle_2015, engel_detection_2015, vodolazov_single-photon_2017, kozorezov_fano_2017}, it is currently understood that most features of photodetection in SNSPDs can be explained by a combination of Fano fluctuations \cite{kozorezov_fano_2017} and vortex-based breaking of superconductivity \cite{vodolazov_vortex-assisted_2015}. More recently, the measurement of record low timing jitter \cite{korzh_demonstrating_2018} has led to a new effort in understanding the latency of SNSPDs \cite{allmaras_2018} in order to predict the intrinsic timing jitter of these detectors. It is known that a precise estimate of the depairing current of a device is needed in order to match experimental results using these models. The most common way of estimating the depairing current is through the Kupryianov-Lukichev formula \cite{kupriyanov_1980}, which requires several independent material parameters such as the diffusion coefficient, sheet resistance, critical temperature and nanowire geometry. In this work, we demonstrate a direct method of accessing the depairing current by measuring the kinetic inductance change as a function of the bias current. This method relies on fitting the theoretical dependence calculated by Clem and Kogan \cite{clem_kinetic_2012} where the depairing current is the single free fitting parameter. Having access to a direct measurement of the depairing current enables a better estimation of the figure of merit for the quality of superconducting nanowires: the constriction factor $C$ \cite{Kerman_Constriction_2007}, ratio of the switching and depairing currents ($C = I_{\text{sw}}/I_{\text{dep}}$), since reaching higher fractions of the depairing current gives rise to higher internal detection efficiency and lower intrinsic jitter~\cite{korzh_demonstrating_2018}.

\quad The kinetic inductance dependence on bias current is determined by measuring the self-resonance of a superconducting nanowire in a coplanar waveguide (CPW) structure, using a vector network analyzer (VNA). The resonances were measured in both transmission and reflection modes by analyzing the complex spectral response. Measurement of the self-resonance has been demonstrated for meandered nanowires~\cite{santavicca_resonance_2016}, however, the change of the kinetic inductance at the highest achievable bias current relative to the zero bias current case was less than 10\%, making it difficult to distinguish whether the experiment falls within the fast or slow relaxation category and giving rise to significantly different depairing current predictions~\cite{clem_kinetic_2012}. Here we demonstrate a kinetic-inductance change as high as 31\% for tungsten silicide (WSi) and 28\% for niobium nitride (NbN) nanowires, which allows us to conclude that the experiment falls into the fast relaxation regime by comparing the quality of the fit of the two models. The improvement could be attributed to several factors such as optimized material~\cite{dane_bias_2017}, lower base temperature and use of a cryogenic bias-tee and amplifier.

\quad We present the dependence of the measured depairing current on the width of the nanowire resonators as well as the operating temperature. An important observation is that $C$ reduces for higher operating temperatures for both polycrystalline NbN and amorphous WSi devices, which has significant consequences for design of high-performance SNSPDs at elevated temperatures. There is also an indication that narrower nanowires achieve a lower $C$, which may point to nanowire edge roughness due to fabrication imperfections.

\section{\label{sec:fab}Device Design and Fabrication}

\begin{figure}
\includegraphics[width=\linewidth]{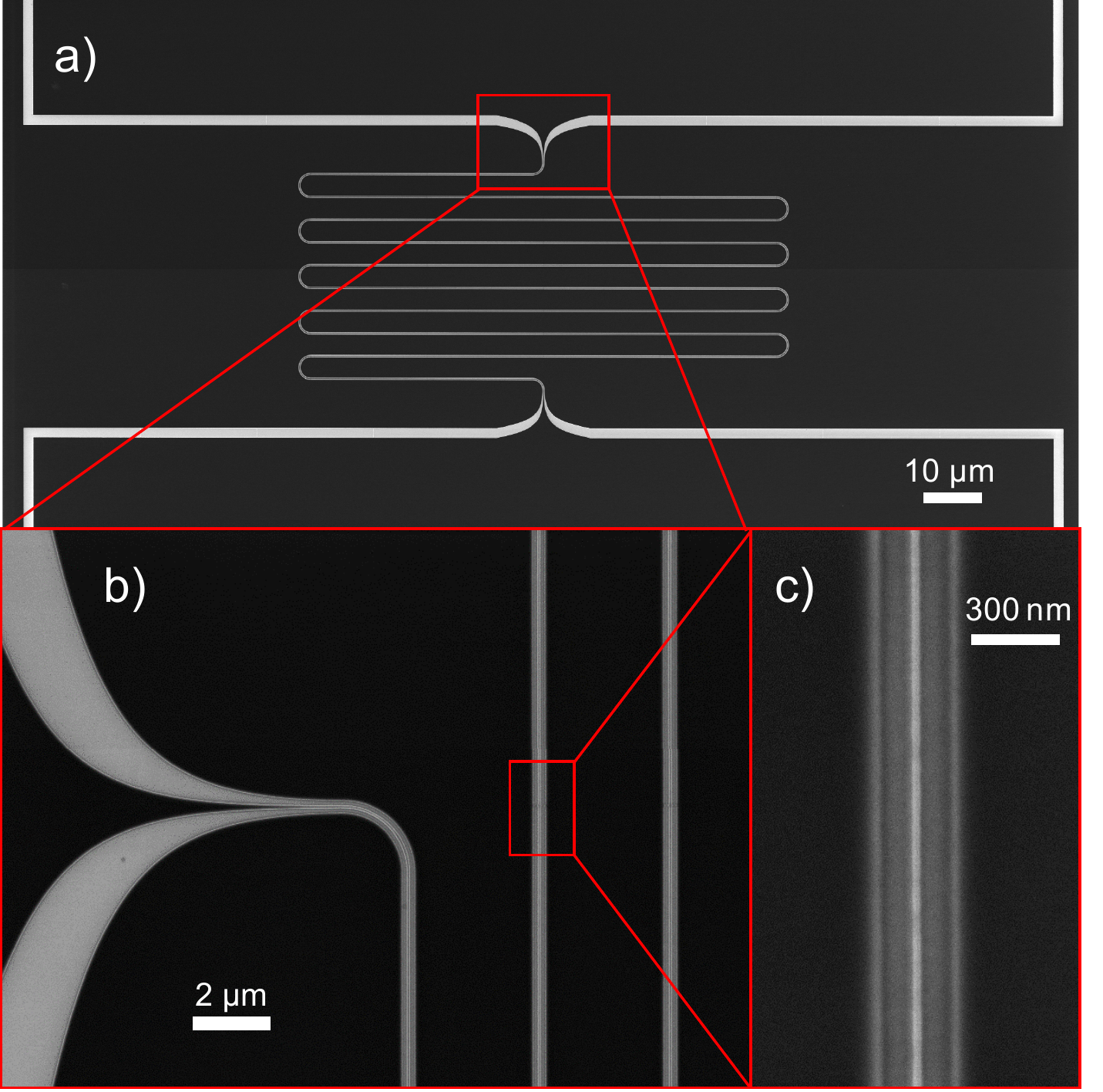} \\
\captionsetup{justification=raggedright, singlelinecheck=false}
\caption{Scanning electron micrographs of NbN CPW resonator used in experiment.  \textsf{\textbf{(a)}} The narrow, meandered nanowire CPW is placed between two wide, 50 $\Omega$ leads (also in CPW configuration), forming a transmission-line resonator. \textsf{\textbf{(b)}} Transition from the 50 $\Omega$ lead to the k$\Omega$ nanowire. \textsf{\textbf{(c)}} Zoomed-in view of the nanowire CPW.}
\label{Fig:SEM}
\end{figure}

\quad The nanowire resonators are designed in a CPW \cite{zhao_single-photon_2017,Zhu_2019} to avoid electromagnetic coupling within the meander and to allow for simplified impedance engineering. The resonance is set up by means of the impedance mismatch between the transmission line and the narrow nanowire. This approach simplifies current biasing of the nanowire. The devices were designed in order to have the resonant frequency at roughly 2 GHz, so that the microwave period ($\tau_{\text{exp}} \approx$ 500~ps) is much larger than the relaxation time of the superconducting order parameter $\tau_{\text{s}}$ for both WSi \cite{Zhang_WSi_2018} and NbN \cite{Zhang_NbN_2018}. An estimate of the relaxation time of the order parameter is given by $\tau_{\text{s}} = \hbar / k_B (T_c-T)$, so for NbN films ($T_c=$~8.65~K) the order parameter relaxation time is 1~ps, while for WSi films ($T_c=$~3.50~K) it is 3.1~ps, at a temperature of 1.05~K. At the highest temperature investigated ($0.8T_c$ for NbN and $0.7T_c$ for WSi) the order parameter relaxation time is 4.6~ps and 7.3~ps, for NbN and WSi films, respectively.

\begin{figure}
\includegraphics[width=\linewidth]{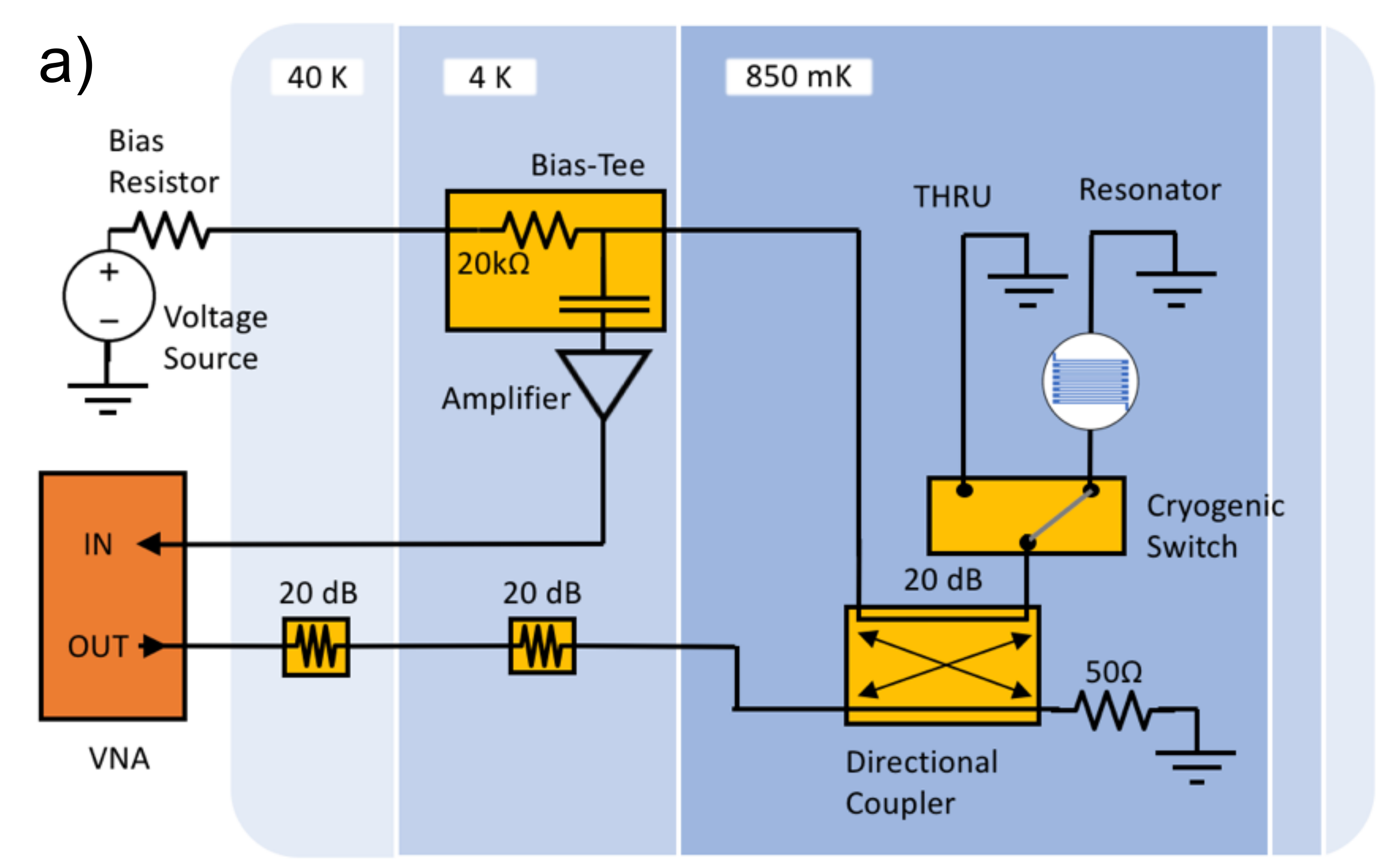} \\
\includegraphics[width=\linewidth]{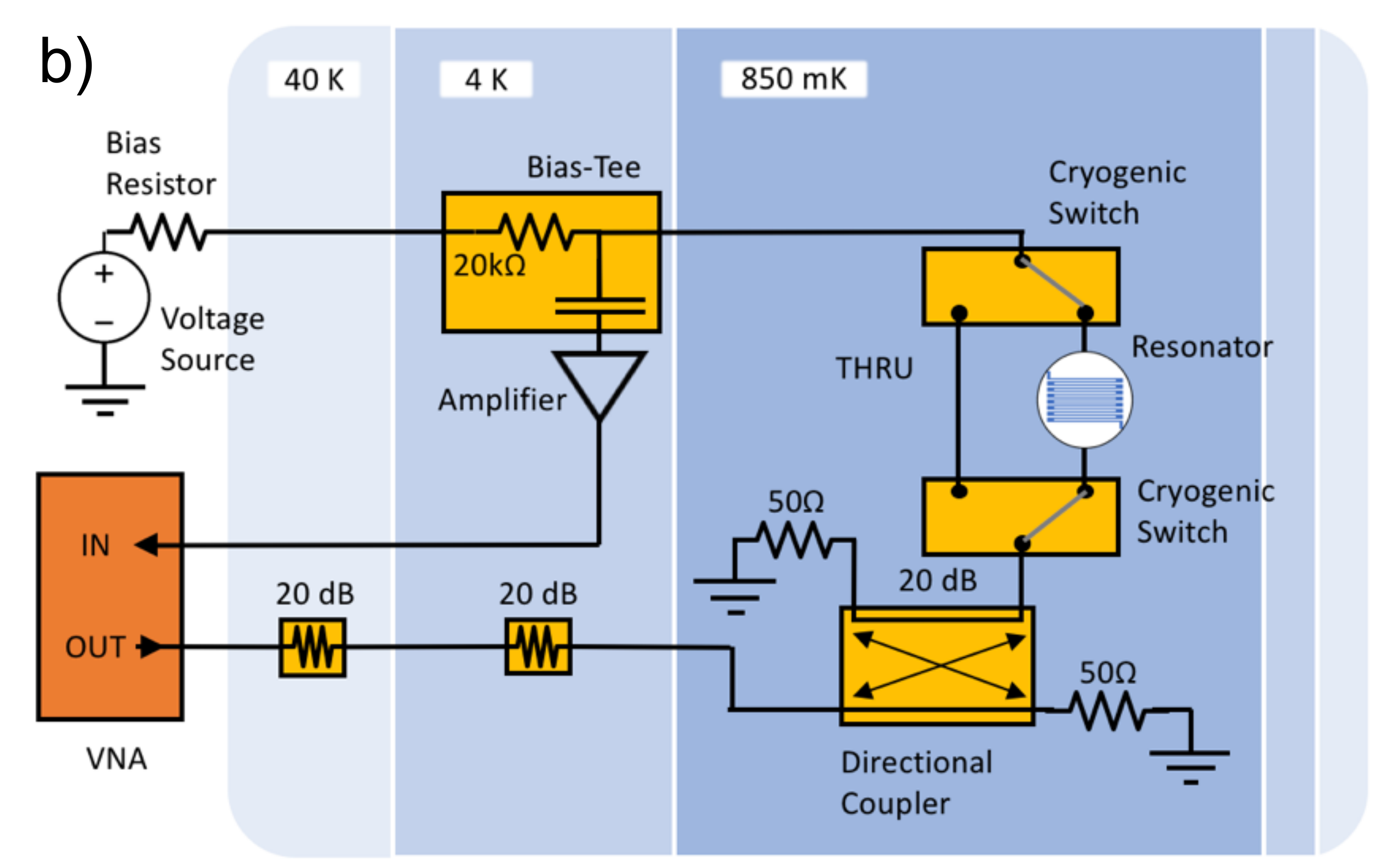}
\captionsetup{justification=raggedright, singlelinecheck=false}
\caption{Schematics for the setup for both the \textsf{\textbf{(a)}} reflection and \textsf{\textbf{(b)}} transmission type measurements. The THRU devices were 50 ohm superconducting CPW fabricated on the same chip used for calibration purposes.}
\label{SetupFig}
\end{figure}

\quad The devices were fabricated from a 6 nm thick NbN film and from a 7 nm thick WSi film. NbN film was sputter deposited on a 4-inch silicon wafer with a 300 nm thick thermal oxide layer \cite{dane_bias_2017}. WSi was sputter deposited on 4-inch silicon wafer with a 240 nm thick thermal oxide layer and was passivated with a 15 nm thick silicon dioxide (SiO$_2$) film. All the devices and pad structures were patterned using $125\,\mathrm{kV}$ electron beam lithography with gL2000 positive tone resist \cite{zhao_single-photon_2017}. The patterns were then transferred into NbN and WSi by CF$_4$ reactive ion etching. A layer of HSQ was spun on the dies after fabrication, for passivation. 

\section{\label{sec:setup}Experimental Setup}

\quad The experimental setup is illustrated in Fig.~\ref{SetupFig}. We measured the resonant frequency of the SNSPD-like resonators (Fig.~\ref{Fig:SEM}) both in transmission mode \cite{santavicca_resonance_2016} and in reflection mode (Fig.~\ref{SetupFig}).
The devices were cooled to a base temperature of 1.05~K with a cryocooler composed of a pulse tube followed by a Helium-4 sorption cooler.

\quad The device resonance was measured with a 300~kHz-6~GHz VNA. The output signal from the VNA was attenuated by 20~dB at both the 40~K and 4~K stages, before entering the input port of a 20~dB directional coupler. The transmission port of the coupler was 50~$\Omega$ terminated, while the coupling port, was connected to an RF switch on the ~1~K stage. One port of the switch was connected to a through device, which consisted of a superconducting CPW used for calibration purposes. The RF switch was used to achieve the same electrical environment between the calibration device and the device under test. 

\quad While measuring in the transmission mode, the output port of the resonator CPW, was connected to a second switch followed by a cryogenic bias tee. The DC port was used to current bias the nanowire, while the RF port was fed to the input of a SiGe cryogenic amplifier (Cosmic Microwave, CITLF1 \footnote{The use of trade names is intended to allow the measurements to be appropriately interpreted and does not imply endorsement by the US government, nor does it imply these are necessarily the best available for the purpose used here.}). The bias tee and amplifier were mounted and thermalized to the 4~K stage. The amplified RF signal was fed to the input port of the VNA. Finally, the isolated port of the directional coupler was 50~$\Omega$ terminated to guarantee current flow through the nanowire. In reflection mode, the nanowire was connected to the coupler on one side and grounded on the other. In this configuration, the isolation port of the directional coupler connected the nanowire to the bias tee and amplifier. For both scenarios, the power output of the VNA was adjusted such that the RMS current flowing through the resonator was of the order of 100 nA, which is small to prevent a shift in the resonant frequency.

\begin{figure*}
\includegraphics[width=.49\linewidth]{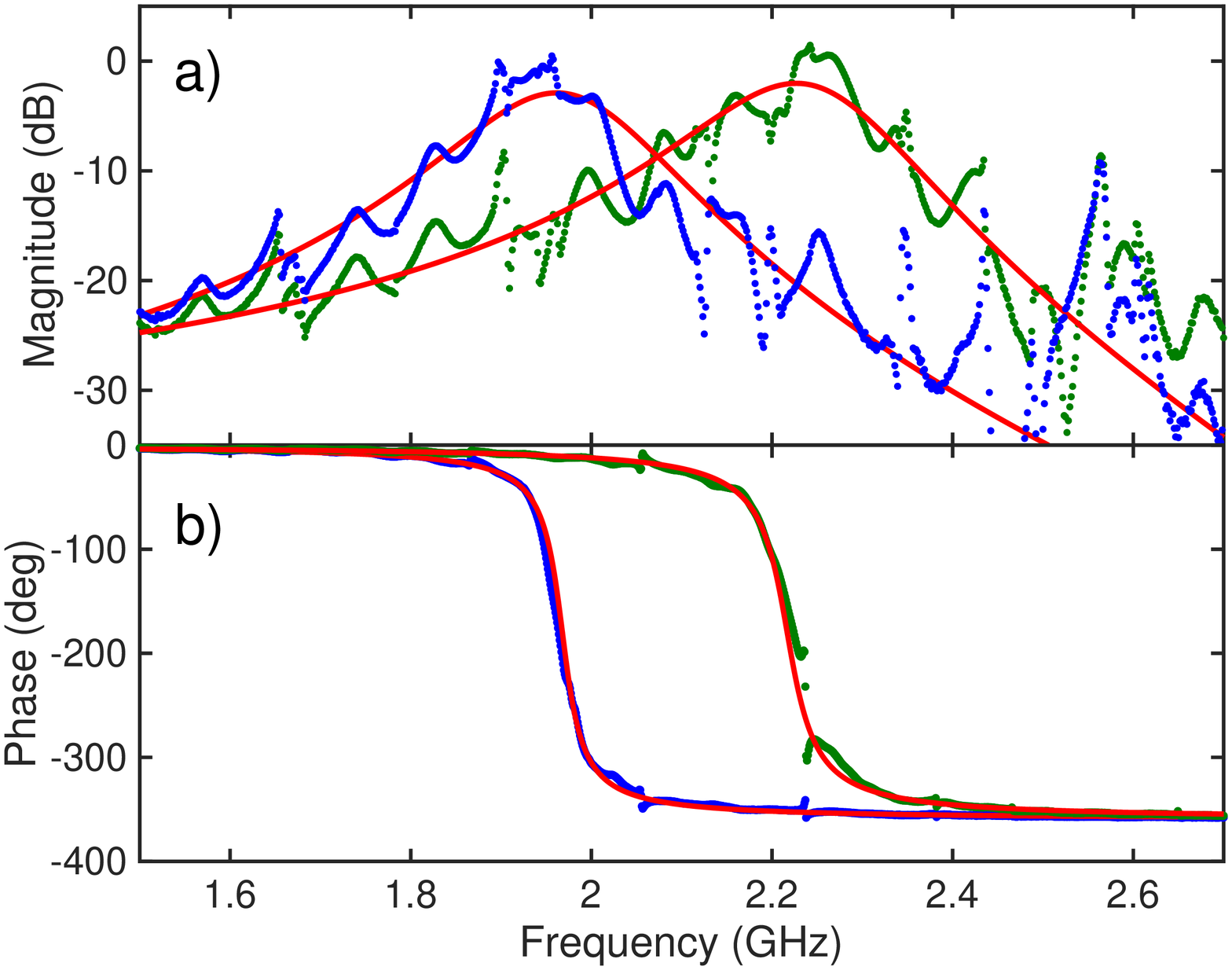}
\includegraphics[width=.49\linewidth]{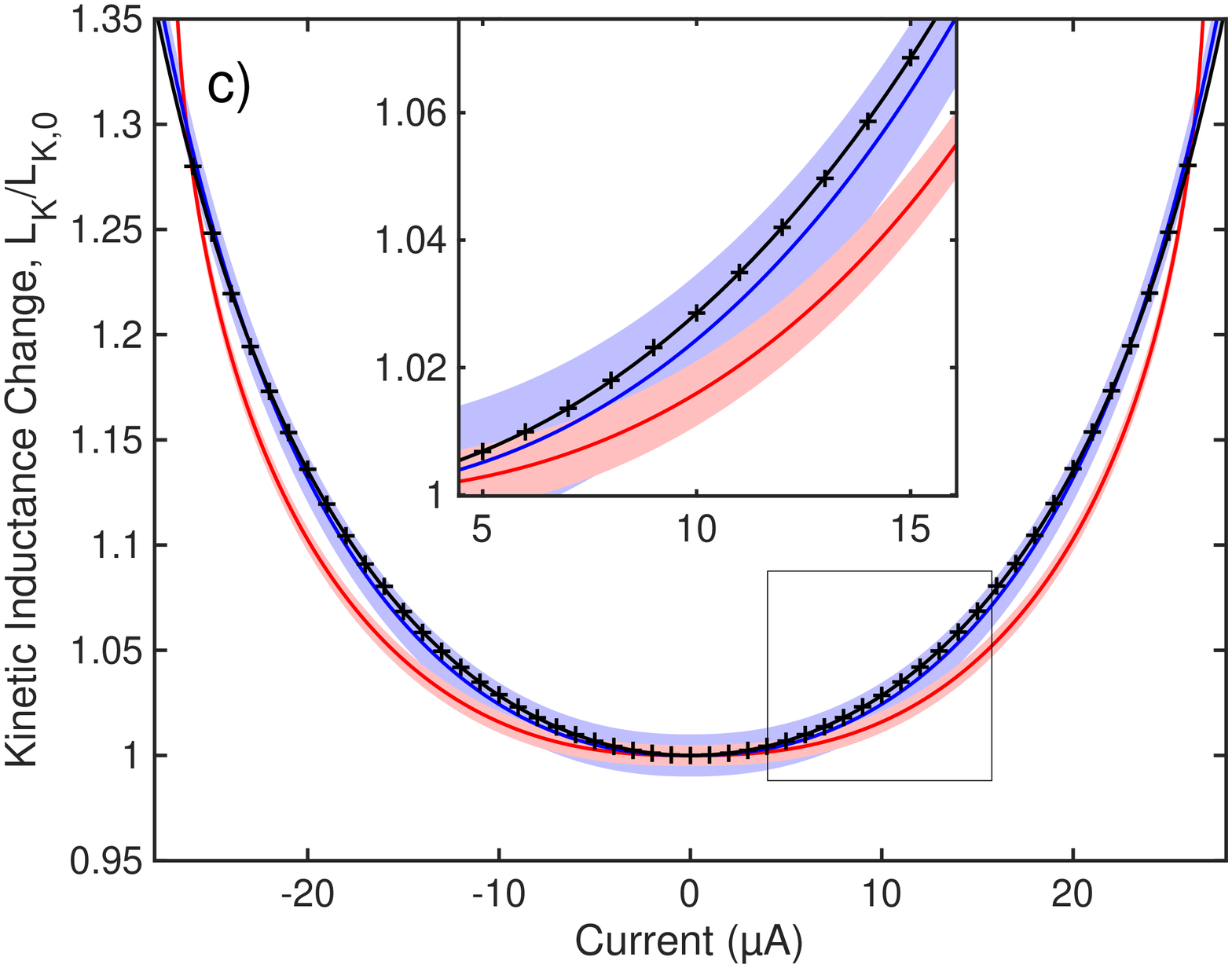}
\captionsetup{justification=raggedright, singlelinecheck=false}
\caption{The measured and the fitting functions for the resonance \textsf{\textbf{(a)}} magnitude and \textsf{\textbf{(b)}} phase responses at zero and near the switching bias current for the 120~nm wide NbN device. 
Fitting of the kinetic inductance ratio of the nanowire \textsf{\textbf{(c)}} using the fast (in blue) and the slow (in red) relaxation approximation models for the 120~nm wide NbN device. The shaded area represents the model's accuracy of 1\% and 0.5\% for the fast and slow relaxation approximation models, respectively, according to Clem and Kogan \cite{clem_kinetic_2012}. In black, the numerical simulation of the kinetic inductance change using the fast relaxation model. It can be shown by the enlarged window that the numerical simulation fits the sample points better than the fast relax approximation model. The estimated depairing current from the models is \SI{38.19}{\micro\ampere} for the fast relaxation approximation, \SI{38.78}{\micro\ampere} for the fast relaxation numerical simulation and \SI{27.05}{\micro\ampere} for the slow relaxation.}
\label{fig:res1}
\end{figure*}

\section{\label{sec:model}Models}

\quad The measured resonance peaks were fitted using a RLC resonator model with a purely reactive bypass channel. For the reflection mode measurement, the resonance was fitted using a double notch filter at the resonant frequency. The magnitude and phase functions of $S_{11}(\omega)$ are written as
\begin{subequations}
\begin{align}
\|S_{11}(\omega)\| &= - I\frac{\big(\frac{1}{2}\Gamma\big)^2}{(\omega - \omega_r)^2 + \big( \frac{1}{2}\Gamma \big) ^2} \label{eq:s11mag}\text{,} \\
\text{arg} \big\{ S_{11}(\omega) \big\} &= -180 + 2\times \tan^{-1}\big[ 2Q (1 - \frac{\omega}{\omega_r}) \big] \label{eq:s11phas}\text{,}
\end{align}
\end{subequations}
where $\Gamma$ is the full-width at half-maximum of the Lorentzian function, $I$ is the peak height, $\omega_r$ is the resonant frequency and $Q$ is the quality factor.

\quad For the transmission mode measurement, the resonance is still been modeled as a Lorentzian function, but accounts for the effect of a bypass channel, modeled as a pure capacitance, in a correction factor. We define $S_{21}(\omega)$ according to
\begin{equation}
\|S_{21}(\omega)\| = I\frac{\big(\frac{1}{2}\Gamma\big)^2}{(\omega - \omega_r)^2 + \big( \frac{1}{2}\Gamma \big) ^2} \Big| 1 - \xi (\omega - \omega_r) \Big| ^2 \text{,}
\label{eq:s21}
\end{equation}
where the correction factor to the Lorentzian function in \eqref{eq:s21} is valid for purely reactive bypass channels and $\xi$ is a constant representing the coupling between the resonator and the reactive channel. For further information regarding the physical meaning of $\xi$, we direct readers to the supplementary information of Weinstein and Schwab \cite{wollman_2014}. 

\quad Once the resonant frequency of the nanowire was evaluated, we could estimate the change in kinetic inductance with increasing bias current according to $\omega_r \propto 1/\sqrt{LC}$ for an RLC resonator. We then fitted the kinetic inductance ratios as obtained using the two relaxation models from Clem and Kogan \cite{clem_kinetic_2012}
\begin{subequations}
\begin{align}
y_{\text{fr}}(x) &= (1 - x^n)^{1/n} \text{,}
\label{eq:fr} \\
y_{\text{sr}}(x) &= y_0 - (y_0 - 1)(1 - x^n)^{1/n} \text{,}
\label{eq:sr}
\end{align}
\end{subequations}
where $y = \mathcal{L}_{\text{k}}(q,t)/\mathcal{L}_{\text{k,0}}(t)$ is the ratio between the kinetic inductance of the biased superconducting nanowire and the kinetic inductance at zero bias current, $y_0$ and $n$ are fixed parameters defined by Clem and Kogan \cite{clem_kinetic_2012} for specific temperature ratios $t = T/T_c$, $x = |j_{\text{s}}| /j_{\text{d}}(t)$ is the ratio between the bias current density and the depairing current density and the subscripts ``fr" and ``sr" stand for ``fast relaxation" and ``slow relaxation" respectively. The difference between the two models is related to the characteristic timescale of variation of $j_{\text{s}}$, the current-biased experiment characteristic time $\tau_{\text{exp}}$, with respect to the relaxation time of the superconductor ($\tau_{\text{s}}$). We refer to fast relaxation if the experimental time constant is much larger than the characteristic superconductor relaxation time, while for slow relaxation, the experimental time constant is much smaller. The accuracy of the fitting functions \eqref{eq:fr} and \eqref{eq:sr} compared to the full numerical solution presented in \cite{clem_kinetic_2012} is 1\% for the fast relaxation model and 0.5\% for the slow relaxation model. As a comparison, we also calculate the depairing current using a fit to the full numerical results of the fast relaxation model using the approach of \cite{clem_kinetic_2012} and keeping 15000 modes in the numerical calculations. The numerical results provide a better match to the experimental results than the approximate equation with only a small change in the extracted depairing current when compared to the approximate fit of (\ref{eq:fr}). Within both models, the depairing current density $j_{\text{d}}(T)$ is the only fitting parameter.

\section{\label{sec:results}Results}

\quad We measured the resonant frequencies of nanowire devices with widths of typical SNSPDs (50-200~nm) using both NbN and WSi thin films. 

\quad The resonance features of an NbN, 120nm wide device, measured at zero current and close to the switching bias current ($\text{I}_{\text{bias}} = \SI{26}{\micro\ampere}, \text{I}_{\text{sw}} = 27 \pm .5 ~\SI{}{\micro\ampere}$) are shown in Fig.~\ref{fig:res1}.  The fit of the models described in section \ref{sec:model} is shown in red for both the transmission (Fig.~\ref{fig:res1}a, where we use \eqref{eq:s21} to fit the magnitude) and reflection (Fig.~\ref{fig:res1}b, where we use \eqref{eq:s11phas} to fit the phase) measurements. For the phase analysis, we found it best to normalize the phase data with respect to the phase of the resonator while in non-superconducting state, i.e. biasing the device above its switching current. The resonant peaks obtained using the two different methods of reflection and transmission match within 0.5\%; however, from the goodness of the two fits, we decided to prioritize the analysis of the phase in reflection method as it is, in general, less noisy and requires fewer free parameters to perform the fit. From this point onward we only refer to data collected from the phase response of the resonator in reflection mode.

\begin{figure*}
\includegraphics[width=.49\linewidth]{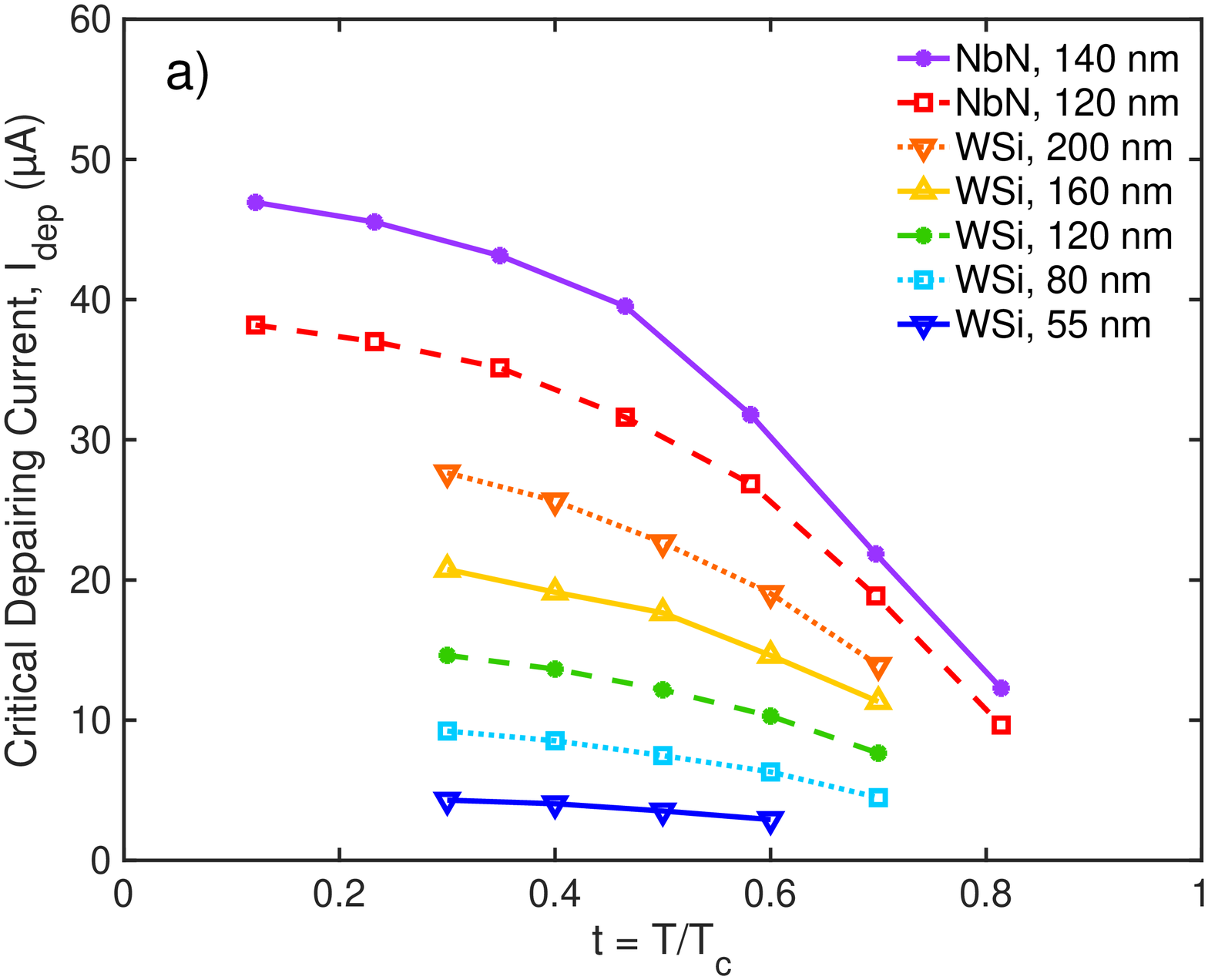}
\includegraphics[width=.49\linewidth]{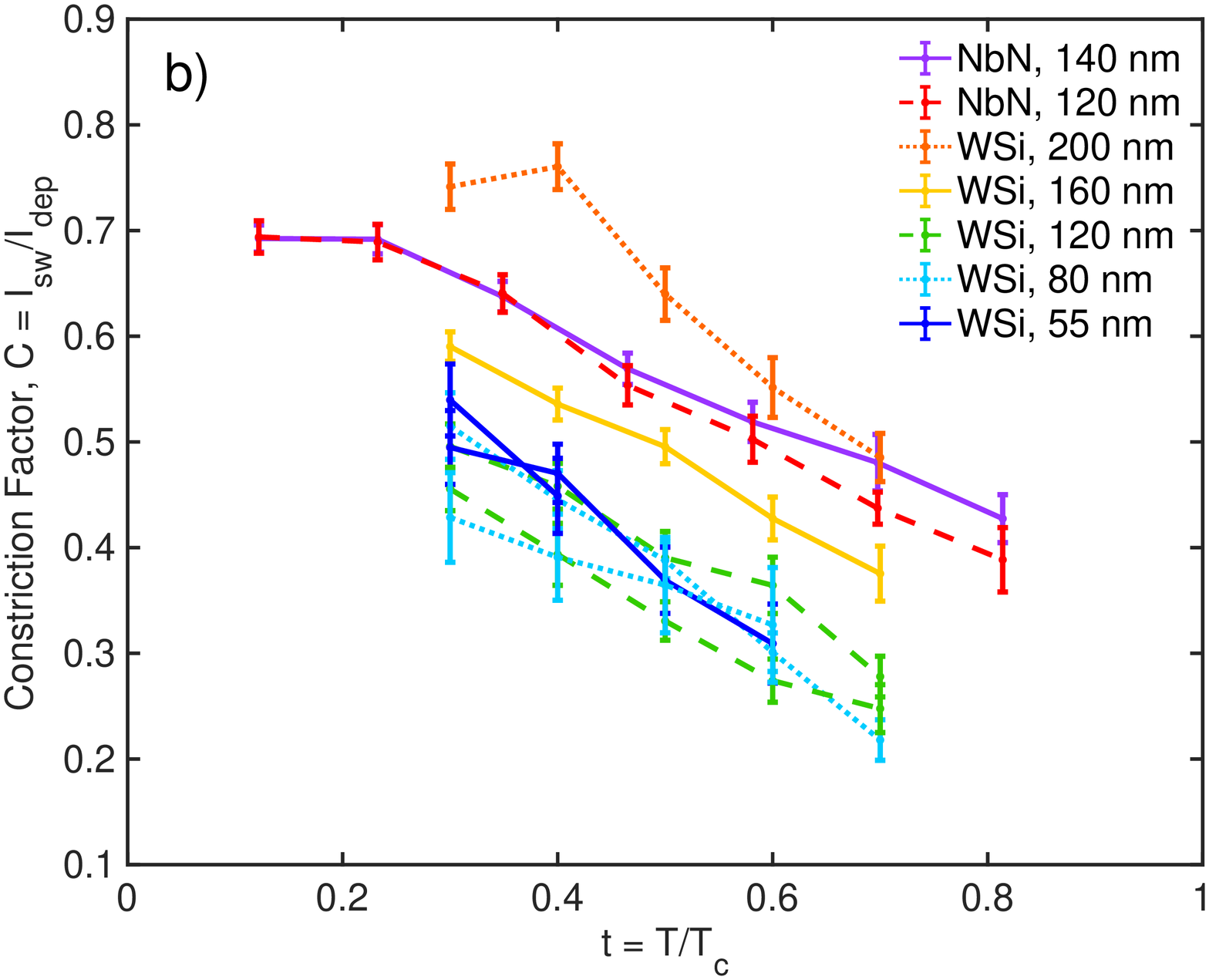}
\captionsetup{justification=raggedright, singlelinecheck=false}
\caption{\textsf{\textbf{(a)}} Estimated depairing current using the fast relax model as a function of the operating temperature for both NbN and WSi devices and \textsf{\textbf{(b)}} the switching to depairing current ratio (constriction factor) for all the tested devices as a function of the fraction of superconductor transition temperature.}
\label{fig:depVsT}
\end{figure*}

The kinetic inductance ratios, obtained by the measured resonance frequencies as,
\begin{equation}
y(\text{I}_{\text{bias}},T) = \frac{\mathcal{L}_{\text{k}}(q,T)}{\mathcal{L}_{\text{k,0}}(T)} = \Bigg[\frac{\omega_r(\text{I}_{\text{bias}} = 0,T)}{\omega_r(\text{I}_{\text{bias}},T)}\Bigg]^2 \text{,}
\end{equation}
are then plotted in Fig.~\ref{fig:res1}c a function of the bias current. The fast relaxation and slow relaxation models discussed by Clem and Kogan \cite{clem_kinetic_2012} have been fitted to the data, where the only free parameter is the depairing current $I_{\text{dep}}$ of the nanowire. The estimated depairing current for each model can be found in the caption of Fig.~\ref{fig:res1}.

\quad It is immediately clear from Fig.~\ref{fig:res1}c that the fast relaxation model provides a better fit of the experimental data than the slow relaxation model. Moreover, the depairing current evaluated using the latter model appears to be unreliable since the model predicts depairing currents just above the measured switching currents. Due to fabrication imperfections, the measured switching currents in SNSPDs are typically significantly below the depairing current since the switching current is set by the weakest point along the nanowire, typically referred to as a constriction. The depairing current, however, is an average characteristic of the nanowire, hence a switching current approaching the depairing current would suggest a "perfect" nanowire. By removing the highest bias current points, it is possible to simulate a more constricted nanowire, while the measured depairing current should remain unchanged. Carrying out this exercise, the slow relaxation model does not predict constant values while the fast relaxation model is robust and provides depairing current estimates which are more consistent with theoretical models. With this, we conclude that our experiment falls into the fast relaxation regime, which has not been confirmed previously~\cite{santavicca_resonance_2016}. 

\quad We measured the resonant frequency of the nanowires with respect to biasing at different temperature conditions. This measurement was done to compare the temperature dependence of the depairing current with the theoretical predictions. The NbN nanowires resonance frequencies were collected starting from the base temperature 1.05~K up to 7.00~K, which is more than 80\% of the superconductor transition temperature, measured to be 8.65~K, while the WSi devices were measured up to 2.45~K, which corresponds to 70\% of their $T_c$ of 3.50~K. The constriction factor drops with increasing temperature for both NbN and WSi devices. This effect might be due to local defects in the nanowire structure: if the weakest constriction in the nanowire has a lower transition temperature then its local switching current would drop faster than the depairing current for the whole nanowire, with increasing operating temperature. This observation deserves future investigation, since it could shed light on the possibility of SNSPD operation at elevated temperatures.

\quad In total, we tested one die with two NbN device geometries (widths of 120 and 140~nm) and two identical dies with five WSi device geometries each (widths of 55, 80, 120, 160 and 200~nm). The measured switching currents and the estimated critical depairing currents based on the fast and the slow relaxation models are collected in Table \ref{tab:dep}. In Fig.~\ref{fig:depVsT}a we report the trend of the devices' critical depairing currents with respect to different temperatures. We estimate the zero temperature depairing current $I_{\text{dep}}(0)$ by fitting the measured temperature dependence of $I_{\text{dep}}(T)$ to the function defining the temperature dependence of the numerical solution to the Usadel equations. These estimated values for $I_{\text{dep}}(0)$ are collected in Table \ref{tab:dep}. For comparison, we also calculated the theoretical critical depairing current at zero temperature according to Kupryianov and Lukichev model \cite{kupriyanov_1980}, denoted as:
\begin{equation}
I_{\text{dep}}^{\text{KL}}(0) = 1.491~e~N(0)~[\Delta(0)]^{3/2}\sqrt{D/\hbar}~w d
\end{equation}
where $e$ is the electron charge, $N(0) = (2e^2DR_\text{sq}d)^{-1}$ is the single-spin electron density of states at Fermi level in the normal state, $\Delta(0) = 1.764~k_BT_c$ is the superconducting gap at zero temperature, $D$ is the diffusion coefficient, $R_\text{sq}$ is the square resistance, and $w$ and $d$ are width and thickness of the nanowire, respectively. In order to calculate these values, we measured the diffusion coefficient for WSi, while for NbN, we used a value found in literature. 

\begin{table*}
\begin{tabularx}{\textwidth}{ccYYYYYYYYcYY}
\hline
 \multicolumn{2}{>{\hsize=\dimexpr2\hsize+2\tabcolsep+\arrayrulewidth\relax}Y}{\textbf{Device}} &  \multicolumn{2}{>{\hsize=\dimexpr2\hsize+2\tabcolsep+\arrayrulewidth\relax}Y}{\textbf{Fast Relax Approximation}} & \multicolumn{2}{>{\hsize=\dimexpr2\hsize+2\tabcolsep+\arrayrulewidth\relax}Y}{\textbf{Fast Relax Numerical}} & \multicolumn{2}{>{\hsize=\dimexpr2\hsize+2\tabcolsep+\arrayrulewidth\relax}Y}{\textbf{Slow Relax Approximation}} & \multicolumn{3}{>{\hsize=\dimexpr3.5\hsize+2\tabcolsep+\arrayrulewidth\relax}Y}{\textbf{Measured}} & \multicolumn{2}{>{\hsize=\dimexpr2\hsize+2\tabcolsep+\arrayrulewidth\relax}Y}{\textbf{Estimated}} \\
Material & Width & $I_{\text{dep}}$ & Fit $R^2$ & $I_{\text{dep}}$ & Fit $R^2$ & $I_{\text{dep}}$ & Fit $R^2$ & $I_{\text{sw}}$ & $C$ & $\Big\{ \frac{\mathcal{L}_{\text{k}}(q,T)}{\mathcal{L}_{\text{k,0}}(T)} \Big\}_{\text{sw}}$ & $I_{\text{dep}}(0)$ & $I_{\text{dep}}^{\text{KL}}(0)$\\[.5\normalbaselineskip] \hline
WSi & 55 nm & 4.40 & 0.9942 & 4.67 & 0.9990 & 3.32 & 0.7413 & 2.25 & 0.54 & 1.107 & 5.31* & 6.47 \\
\rowcolor{gray!10}WSi & 55 nm & 4.29 & 0.9924 & 4.59 & 0.9984 & 3.09 & 0.7998 & 2.13 & 0.49 & 1.085 & 5.09 & 6.47 \\
WSi & 80 nm & 7.58 & 0.9808 & 8.30 & 0.9962 & 4.74 & 0.9636 & 3.25 & 0.43 & 1.055 & 9.72 & 10.68 \\
\rowcolor{gray!10}WSi & 80 nm & 9.22 & 0.9955 & 9.81 & 0.9995 & 6.05 & 0.9838 & 4.75 & 0.52 & 1.094 & 11.66 & 10.68 \\
WSi & 120 nm & 14.62 & 0.9930 & 15.60 & 0.9996 & 9.47 & 0.9801 & 7.25 & 0.50 & 1.090 & 18.72 & 17.41 \\
\rowcolor{gray!10}WSi & 120 nm & 14.82 & 0.9940 & 16.25 & 0.9961 & 9.55 & 0.9834 & 6.75 & 0.46 & 1.066 & 20.31 & 17.41 \\
WSi & 160 nm & 20.76 & 0.9980 & 21.71 & 0.9986 & 13.82 & 0.9856 & 12.25 & 0.59 & 1.152 & 26.07 & 23.46 \\
\rowcolor{gray!10}WSi & 160 nm & 21.16 & 0.9970 & 21.98 & 0.9984 & 14.44 & 0.9750 & 13.50 & 0.64 & 1.179 & 25.98 & 23.46 \\
WSi & 200 nm & 27.65 & 0.9954 & 28.06 & 0.9993 & 21.00 & 0.7813 & 20.50 & 0.74 & 1.313 & 33.26 & 30.28 \\
\rowcolor{gray!10}NbN & 120 nm & 38.19 & 0.9975 & 38.78 & 1.0000 & 27.05 & 0.9239 & 26.50 & 0.69 & 1.280 & 42.07 & 43.30\\
NbN & 140 nm & 46.93 & 0.9970 & 47.67 & 0.9999 & 33.09 & 0.9295 & 32.50 & 0.69 & 1.280 & 51.47 & 50.52\\ \hline
\end{tabularx}
\captionsetup{justification=raggedright, singlelinecheck=false}
\caption{Table representing the results obtained at base temperature (1.05~K). (*) The estimated depairing current at zero Kelvin for the 55~nm wide nanowire is estimated fitting only two points. Switching currents ($I_{\text{sw}}$) were extracted from IV curves, measured at a rate of several minutes per sweep.}
\label{tab:dep}
\end{table*}

\quad In order to calculate these values, we measured the temperature dependence of the upper critical magnetic field ($B_{c2}$) and extracted information on material properties of the WSi thin film.  The electron diffusion coefficient $D$ was obtained from the slope of the $B_{c2}$ vs $T$ curve. In the limit of a dirty superconductor, the electron diffusivity $D$ can be expressed as follows, based on \cite{Semenov_2009},
\begin{equation}
D = \frac{1.097}{\Big[-\frac{dB_{c2}(T)}{dT}\Big]_{T=T_c}}, \label{eq:D}
\end{equation}
where the diffusion coefficient $D$ has dimensions of $[\text{cm}^2\text{s}^{-1}]$, the upper critical magnetic field $B_{c2}$ has dimensions of [T] and the temperature has dimensions of [K].

\quad The external magnetic field was applied perpendicular to the surface of the film and $B_{c2}$ was defined as the field at which the resistance of the film becomes half of the normal state value. The calculated value for the electron diffusion coefficient, based on equation \eqref{eq:D}, is 0.74 $\text{cm}^2/\text{s}$ for the 7~nm thick WSi film. The Ginzburg-Landau coherence length, $\xi_{GL}(0)$, at $T=0$ can be extracted from the following equation:
\begin{equation}
B_{c2}(T) = \frac{\Phi_0}{2~\pi~\xi(T)^2}, \label{eq:BC2}
\end{equation}
where $\Phi_0 = h/2e$ is the magnetic-flux quantum and $e$ is the electron charge.

\quad In the limit of a dirty superconductor, a linear extrapolation of the measured $B_{c2}(T)$ down to $T = 0$, overestimates the real upper critical field at zero temperature and consequently underestimates the superconducting coherence length.  A more realistic value of $B_{c2}(0)$ is given by
\begin{equation}
B_{c2}(0) = 0.69~T_c~\Big[-\frac{dB_{c2}(T)}{dT}\Big]_{T=T_c}.
\end{equation}

\quad Using this value of $B_{c2}(0)$ in equation \eqref{eq:BC2} the calculated coherence length is 9.62 nm for the 7 nm thick WSi film. For the 6~nm thick NbN film, we considered a diffusion coefficient of $D = 0.5~\text{cm}^2/\text{s}$ as used by Zhao \textit{et al.} \cite{zhao_single-photon_2017} and estimated a coherence length at zero temperature $\xi_{GL}(0)$ of 5.01 nm.

\quad We measured the constriction factor, $C(T) = I_{\text{sw}}(T) / I_{\text{dep}}(T)$, which is the ratio between the switching and depairing current, at different temperature conditions for all the devices tested. This ratio can be considered as the quality of the nanowire itself. Shown in Fig.~\ref{fig:depVsT}b, the ratio of currents suggests a decrease of quality of the devices with increasing temperature.

\begin{figure}
\includegraphics[width=\linewidth]{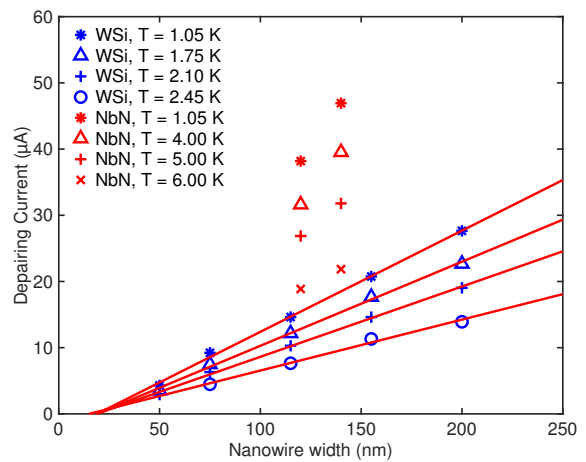}
\captionsetup{justification=raggedright, singlelinecheck=false}
\caption{Depairing current at different temperatures with respect to increasing resonator width for NbN and WSi devices. The linear fit for the WSi devices shows the presence of an offset.}
\label{fig:Res2}
\end{figure}

\quad For the WSi devices, since five geometries were studied, we were able to show the dependence of the depairing current on the device width. As Fig.~\ref{fig:Res2} shows a linear fit to the depairing current estimated at different temperatures seems to suggest that the effective widths of the nanowires might be reduced from the measured widths (SEM after etching)  by an offset of $\sim 23$~nm. That effect could be caused by the loss of superconductivity in the edges of the nanowire due to scattering of particles during etching, or due to oxidation of the nanowire caused by exposure to the environment. It is worth noting that the offset is close to two times the superconducting coherence length $\xi_{GL}$ of the WSi devices, so it is possible that poisoning of the edges of the nanowire during the fabrication process might have suppressed the superconducting active area by roughly one coherence length on each side. More work is needed to conclusively determine the the cause of this observation.

\section{\label{sec:conc}Conclusion}

\quad We have demonstrated a reproducible experimental setup able to estimate the depairing current of superconducting nanowires. According to our experimental data obtained by measuring both NbN and WSi nanowire resonators, the fast relaxation model discussed in the literature gives a more robust and reliable estimate of the depairing current. This experimental method, when combined with other device performance metrics such as the internal efficiency, can be used to refine detection mechanism models and improve the current understanding of the device physics of SNSPDs.

\quad A direct estimation of the depairing current is essential in experimental tests of the relation between the device's minimum photon energy sensitivity and the width of the SNSPD, which is a crucial aspect to take into account when designing SNSPD for specific wavelengths.

\quad Finally, we introduced a new way to measure the quality of the devices in terms of the constriction factor $C$, and we showed that this factor decreases with increasing temperature. The reason for this decrease would make an interesting topic for future study.

\section*{Acknowledgements}

\quad Part of this research was performed at the Jet Propulsion Laboratory, California Institute of Technology, under contract with the National Aeronautics and Space Administration. Support for this work was provided in part by the DARPA Defense Sciences Office, through the DETECT program. J. P. A. acknowledges partial support from the NASA Space Technology Research Fellowship program. D. Z. acknowledges support from the A*STAR National Science Scholarship.

\bibliographystyle{aipnum4-1}
\bibliography{Bibliography.bib}

\end{document}